# Mode-Specific Dynamics of Ammonia Dissociative Chemisorption on Ru(0001)

Xixi Hu,[1,2,*] Minghui Yang,[3] Daiqian Xie,[1,*] and Hua Guo[2,*]

[1]*Institute of Theoretical and Computational Chemistry, Key Laboratory of Mesoscopic Chemistry, School of Chemistry and Chemical Engineering, Nanjing University, Nanjing 210093, China*

[2]*Department of Chemistry and Chemical Biology, University of New Mexico, Albuquerque, New Mexico 87131, USA*

[3]*Key Laboratory of Magnetic Resonance in Biological Systems, National Center for Magnetic Resonance in Wuhan, State Key Laboratory of Magnetic Resonance and Atomic and Molecular Physics, Wuhan Institute of Physics and Mathematics, Chinese Academy of Sciences, Wuhan 430071, China*

*: corresponding authors: xxhu@nju.edu.cn , dqxie@nju.edu.cn, hguo@unm.edu

**Abstract**

The mode specific dissociative chemisorption dynamics of ammonia on the Ru(0001) surface is investigated using a quasi-classical trajectory (QCT) method on a new global potential energy surface (PES) with twelve dimensions. The PES is constructed by fitting 92524 density functional theory points using the permutation invariant polynomial-neural network method, which rigorously enforces the permutation symmetry of the three hydrogen atoms as well as the surface periodicity. The PES enables highly efficient QCT simulations as well as future quantum dynamical studies of the scattering/dissociation dynamics. The QCT calculations yield satisfactory agreement with experiment and suggest strong mode specificity, in general agreement with the predictions of the Sudden Vector Projection model.

Catalysis on metal surfaces is often initiated by dissociative chemisorption (DC) of reactant molecules, which can also be the rate-determining step.[1] In direct activated DC processes, the cleavage of an intra-molecular bond is facilitated by collision of the impinging molecule with the surface, which overcomes a dissociation barrier, leading to two adsorbed fragments. A clear understanding of the DC process is vital for optimizing reaction conditions and for designing new and more effective catalysts. Recently, there has been increasing interest in the understanding of DC dynamics at a quantum state resolved level, which provides detailed information on some important questions, such as the relative efficacy of the translational vs. internal energy of the impinging molecules in promoting the DC reaction.[2-4] Experimental studies with molecular beams and high vacuum have revealed that direct DC is often mode specific, that is, the efficacy in promoting the reaction depends on which mode of the molecule is excited. The mode specificity suggests that dynamics is of paramount importance in direct DC.

Theoretical studies of DC dynamics have revealed, among other things,[5,6] the importance of the dissociation transition state in controlling the mode specificity.[7,8] A useful perspective is provided by the Sudden Vector Projection (SVP) model,[9] in which the ability of a molecular mode, whether translational, rotational, or vibrational, to promote the reaction is attributed to its coupling with the reaction coordinate (RC) at the transition state. The coupling is further approximated in the sudden limit by the projection of the corresponding normal mode vector ($Q_i$) onto the RC vector ($Q_{RC}$): $\eta_i = Q_i \cdot Q_{RC} \in [0,1]$. The SVP model, which can be viewed as a generalization of the venerable Polanyi rules,[10] has successfully explained mode specificity of several DC systems, particularly those with more than one vibrational mode.[8]

To quantitatively compute the dissociation probability, one needs to carry out explicit dynamics calculations. There are two approaches to DC dynamics. One is to compute the potential and forces on the fly from density functional theory (DFT),[11,12] while the other constructs an analytical potential energy surface (PES) instead.[13,14] The former, which is often called Ab Initio Molecular Dynamics (AIMD),[15] is ideal for exploring mechanisms, but it is not efficient, particularly for rare events, because of the expensive on-the-fly force calculations. It is also not amenable to quantum dynamics calculations, which may be necessary to account for quantum effects such as tunneling. The latter approach constructs a Born-Oppenheimer PES by fitting a large number of geometries and energies calculated by DFT. Despite the heavy investment upfront, the availability of a global analytical PES not only increases greatly the efficiency for trajectory based methods for exploring the dynamics, but also permits quantum dynamics calculations, which are necessarily non-local.[8] This approach has been quite successful for constructing the DC PESs for diatoms,[16,17] as well as polyatomics such as water,[18-22] methane,[23-28] and $CO_2$.[29]

In this Letter, we explore the DC dynamics for an important polyatomic system, namely ammonia ($NH_3$). This process is important in ammonia synthesis at high temperature and pressure. In addition, ammonia decomposition has been proposed as an effective means for on-site generation of hydrogen without $CO_x$ byproducts,[30-32] which are poisonous for cathodes in proton exchange membrane fuel cells.[33] The DC of ammonia represents the initial step of the hydrogen production, thus of great importance. Experimental studies have suggested Ru is one of the most effective catalysts for both ammonia synthesis and decomposition.[30-32] DFT calculations have been carried on various Ru surfaces to explore the reaction mechanism.[34-39] The DC of ammonia also offers an excellent proofing ground for understanding mode specificity in DC dynamics of small molecules, as it contains three local N-H bonds similar to the O-H and C-H bonds in isoelectronic $H_2O$ and $CH_4$.[40] However, $NH_3$ differs from the other hydride molecules due to the existence of a large-amplitude umbrella vibration which supports tunneling. Importantly, the dynamics of $NH_3$ DC has been investigated experimentally by Luntz and coworkers on Ru(0001) using a molecular beam apparatus.[41] To gain a better understanding of this important system, we in this Letter report the first global PES for the DC of ammonia on rigid Ru(0001) in full twelve dimensions (12D) based on a large number of DFT points . The DC dynamics of both the ground and vibrationally excited $NH_3$ is investigated for the first time using a quasi-classical trajectory (QCT) method on the new PES.

Spin-polarized plane-wave DFT calculations were performed using the Vienna Ab initio Simulation Package (VASP).[42,43] The Ru(0001) surface was modeled by a 3×3 supercell (1/9 ML) and three-layer slab with the atoms kept at their equilibrium positions. The vacuum space between periodic slabs in the Z direction was set to 14 Å. The kinetic energy cutoff for the plane-wave basis for the valence electrons is 450 eV and core electrons were represented by the projector-augmented wave (PAW) method.[44] The first Brillouin zone was sampled using a Γ-centered 3×3×1 *k*-point grid. A Fermi smearing with a width parameter of 0.1 eV was applied to speed up convergence. The Perdew-Burke-Ernzerhof (PBE) functional[45] was used in total energy calculations. The optimized lattice constants for the Ru are $a=b=2.7251$ Å and $c=4.3067$ Å, which are very close to the experimental values of $a=b=2.7058$ Å and $c=4.2816$ Å.[46] This setup converges the energy to within 0.01 eV.

The reaction path was determined by means of the climbing-image nudged elastic band (CI-NEB) method,[47] using eight images between the initial state (IS, adsorbed $NH_3$) and final state (FS, co-adsorbed $NH_2$ and H). The corresponding transition state (TS) was identified as the image with highest energy along the reaction path and validated by its sole imaginary frequency. The relative energies and geometries of the IS, FS and TS of the CI-NEB calculation are presented in Figs. S1 and S2 in Supporting Information (SI).

The 12D global PES for $NH_3$ dissociation on rigid Ru(0001) was constructed using the permutation invariant polynomial-neural network (PIP-NN) approach,[50,51] with the permutation symmetry of the three H atoms as well as the translational periodicity of the Ru(0001) surface taken into account in the fitting, The PIP-NN training was performed using the Levenberg-Marquardt algorithm by dividing the data set into the training (90%), testing (5%) and validation (5%) sets. The "early stop" method was used to avoid overfitting. Following our previous work,[29,52] a primitive PES was first fit with the points extracted from ~2000 AIMD trajectories. Once the primitive

PES was constructed, more points were sampled by QCT calculations. These points were added to the data set based on certain geometric and energetic criteria, and then used to update the PES. The energetic criterion accepts a point if its energy root mean square errors (RMSEs) among the three fits are larger than 0.1 eV and the average energy of the three fits is less than 5 eV related to the global minimum. For the geometric criterion, the Euclidean distances from the selected point to all existing points in the data set are larger than 0.1 Å to avoid over-sampling. This procedure was iterated until the dissociation probability was converged.

With the aforementioned protocol, a total of 92524 DFT energy points was finally used to construct the PES. This data set includes ~2000 additional points for isolated $NH_3$ in order to give a proper description of the vibrationally excited states. In the PIP-NN fitting, the PES is expressed by an NN with two hidden layers, each with 50 neurons. The input layer of the NN consists of 61 symmetry functions, the forms of which are given in SI. The final PES, which is an average one over the three best fits, has an RMSE of 33.7 meV. As shown in Fig. S3 of SI.

Contour plots of the PES are displayed in Fig. 1 as a function of the N atom height $Z_N$ and one N-H bond length $R_{N-H}$ with the N atom fixed on the specific high symmetry sites of the surface. There exist an adsorption well and a saddle point in each panel. The adsorption well at the top site has the deepest depth of -0.85 eV relative to the asymptote, which is in good agreement with the DFT result. Also the barrier on the top site is the lowest (0.51 eV above the asymptote), which is consistent with the CI-NEB calculation. The N-H distances at the site-specific dissociation barriers are all elongated significantly, suggesting a late-barrier reaction. Although the hcp and fcc sites are distinguishable in our symmetry functions, the PESs at the two sites are quite similar.

The QCT calculations were carried out using VENUS,[53] modified for studying DC dynamics.[54] The $NH_3$ molecule was initially set at 7.0 Å above the metal surface with a momentum along the surface normal. Its initial coordinates and momenta were determined with standard normal mode sampling, assuming quantized energies in the internal degrees of freedom. The umbrella mode of $NH_3$ was assumed to localize in a single minimum and the tunneling between the two equivalent wells was ignored. The propagation time step was 0.10 fs, and the total propagation time was 10 ps. The trajectories were counted as reactive ones when the N-H bond reached 2.5 Å (2.7 Å was tested and no difference was found). Otherwise, the trajectories in which the $NH_3$ was scattered back with 6.0 Å above the surface were counted as nonreactive ones. The reaction probability is given by the ratio between the numbers of reactive and all trajectories. Despite the relatively deep pre-transition state adsorption well, most trajectories are direct except for the lowest energies, where some trapping takes place.

The calculated dissociation probabilities for $NH_3$ in its ground vibrational state on Ru(0001) is compared with the experimental data of Luntz and coworkers in Fig. 2. In qualitative agreement with the experiment, the calculated dissociation probability increases with the translational energy. The agreement between theory and experiment is satisfactory, but not quantitative, most likely due to the rigid surface approximation used here. Indeed, the strong influence of the surface temperature in the experimental data suggests an important role of surface motion.

The effects of vibrational excitations were also investigated using the same method and the results are shown in the Fig. 3. Excitations in all vibrational modes are found to promote the DC reaction at given collision energies. In particular, the doubly degenerate asymmetric stretch ($v_3$) and the symmetry stretch ($v_1$) are the most effective in promoting the reaction, while. On the other hand, the doubly degenerate bending mode ($v_4$) and the umbrella ($v_2$) are much less potent in promoting the reaction. However, as a function of total energy, the dissociation probabilities are very close, which suggesting that increasing energy in the translational mode has roughly the same effect as in the vibrational modes in promoting the $NH_3$ dissociation. This situation is quite similar to the dissociation of $CH_4$, but differs from the DC of $H_2O$ where vibrations are much more effective than translation.[2,4,6,8]

It must be pointed out that the observed mode specificity is not consistent with Polanyi's rules, which predict a high vibrational efficacy for vibration in reactions with a late barrier. Furthermore, these rules do not provide guidance for different vibrational modes in their capacity in enhancing reactivity. To better understand the mode specificity in this system, we turn to the SVP model.[9] The symmetric stretching mode ($v_1$) of NH3 give the largest projection of ($\eta_1 = 0.58$) onto the reaction coordinate at the TS, which is only slightly larger than that of the asymmetric stretch ($v_3$) mode ($\eta_3 = 0.53$). While the bending and umbrella modes have very small overlaps, which are $\eta_4 = 0.07$ and $\eta_2 = 0.05$, respectively. These SVP predictions are consistent with the QCT results on the relative efficacy of the vibrational modes. Besides, the projection of the translational mode is 0.15. The prediction that the N-H stretching vibrations is more effective in promoting the DC reaction than translational energy is not borne out by the QCT results. However, one should bear in mind that the QCT treatment of the dynamics is only approximate because of the complete neglect of quantum effects such as tunneling. In addition, QCT is known to be prone to unphysically strong intramolecular vibrational energy redistribution (IVR), which leaks energy from the excited modes to the rest of the molecule. Indeed, QCT is known to underestimate the effect of vibrational excitation,[21] presumably due to excess IVR. A more accurate characterization of the DC dynamics should rely on quantum dynamical calculations.[8] Indeed, ammonia presents a challenge for quantum dynamics because of the large-amplitude umbrella mode, which also supports tunneling. Work in this direction is already underway in our groups.

It should be further pointed out that although the QCT results reported here should provide useful guidance on the DC dynamics, the calculated dissociation probabilities cannot be directly compared with experimental sticking probabilities. This is because the rigidity of the Ru(0001) surface assumed in our model prevents energy dissipation into surface phonons. Such effects can be quite significant in DC dynamics involving molecules with non-hydrogen atoms.[6] To properly treat energy exchange with surface phonons, the PES needs be extended to include the surface atoms, which is far more challenging but has recent become possible for DC systems.[55-57]

Simulations involving the surface phonons will be necessary to understand the experimentally observed precursor mechanism, which apparently coexists with the direct activation channel.[41] In addition, the surface electron-hole pairs have also been ignored in our calculations, although our recent studies suggested a relatively minor effect for direct DC of molecules.[58,59]

To summarize, an accurate global PES for ammonia DC on Ru(0001) is derived from fitting a large number of DFT points. This analytical PES not only allows the QCT calculations reported here with vastly higher efficiency than AIMD, but it also facilitates future quantum dynamical calculations on both scattering and dissociation. Strong mode specificity has been observed the DC dynamics of $NH_3$, similar to the DC of $CH_4$ and $H_2O$. In particular, the N-H stretching vibrations strongly promote the reaction, while the bending and umbrella modes are substantially less effective in this respect. As a result, it can be concluded that mode specificity, and the associated bond selectivity, is a quite general phenomenon in molecular DC. These theoretical predictions should motivate future experiments on this system.

**Acknowledgements**: This work is supported by the Ministry of Science and Technology (2017YFA0206501) and the National Natural Science Foundation of China (Grant No. 91641104, 21590802, 21733006, and 21773297) and by United States National Science Foundation (CHE-1462019 to H.G.). We thank Prof. Bin Jiang for several useful discussions.

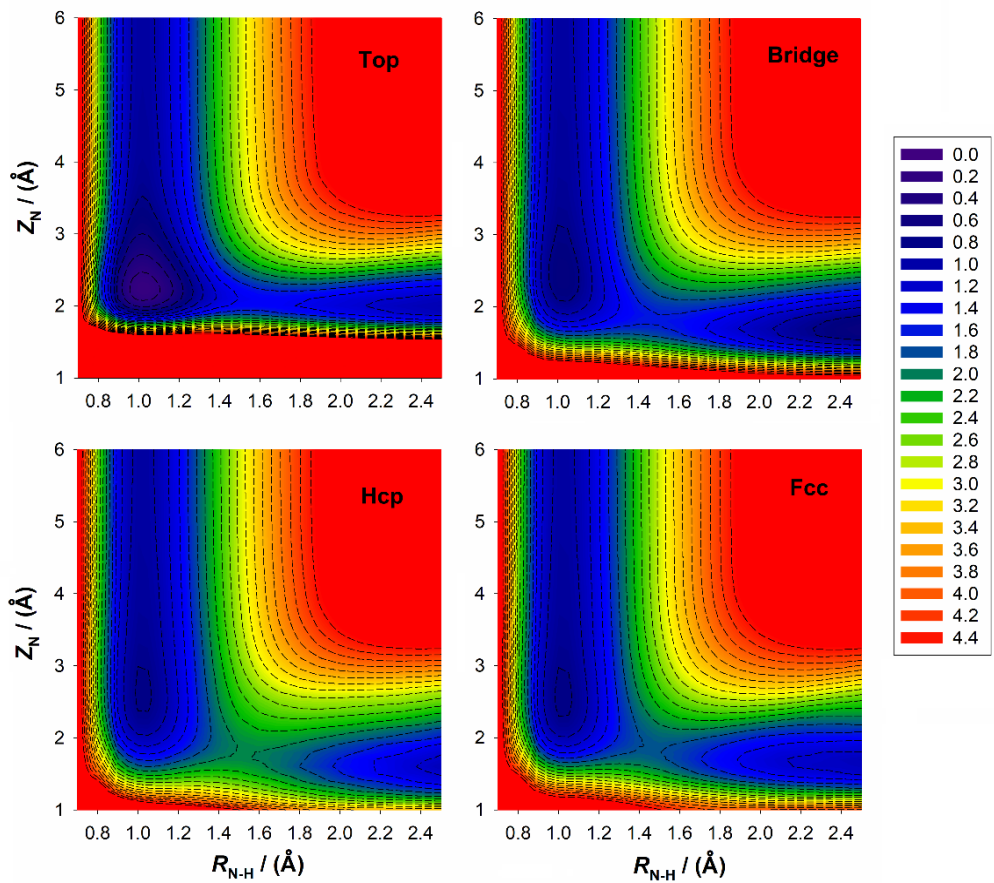

Fig. 1 Contour plots of the $NH_3$+Ru(0001) PES when the N atom are fixed on the top, bridge, hcp and fcc sites, with other coordinates relaxed.

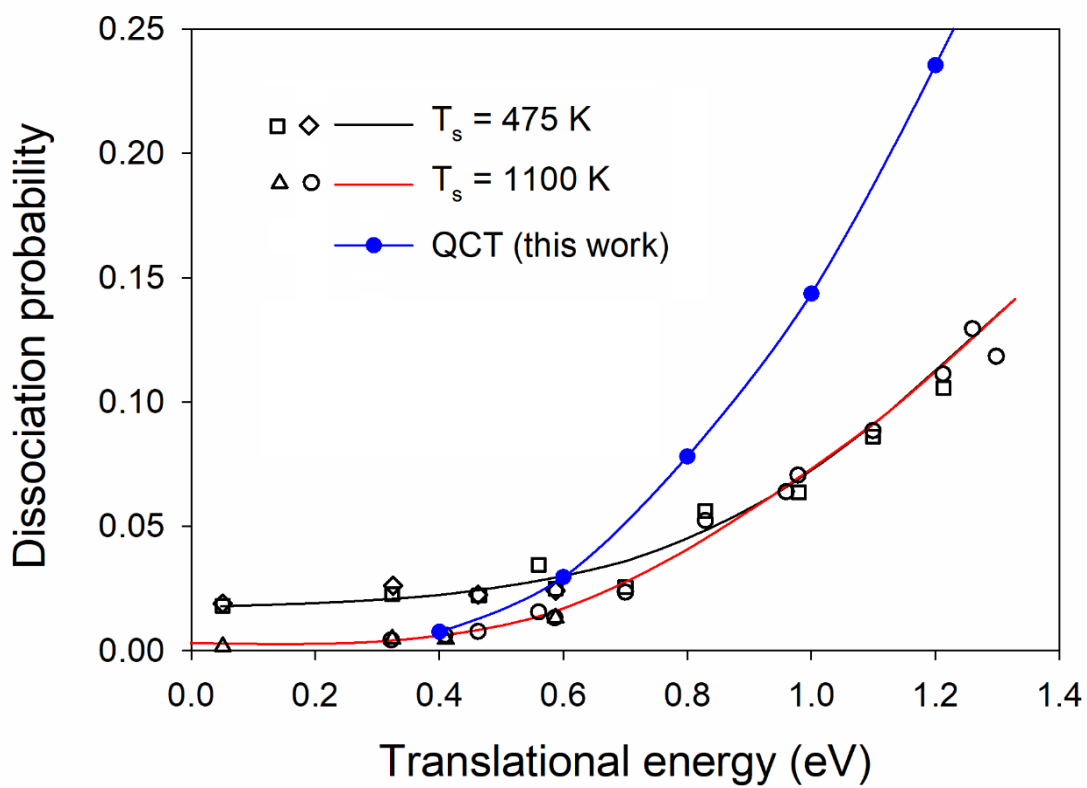

Fig. 2 Comparison of the calculated dissociation probability with the experimental initial sticking coefficients at different surface temperatures ($T_s$) measured by Luntz and coworkers.[41]

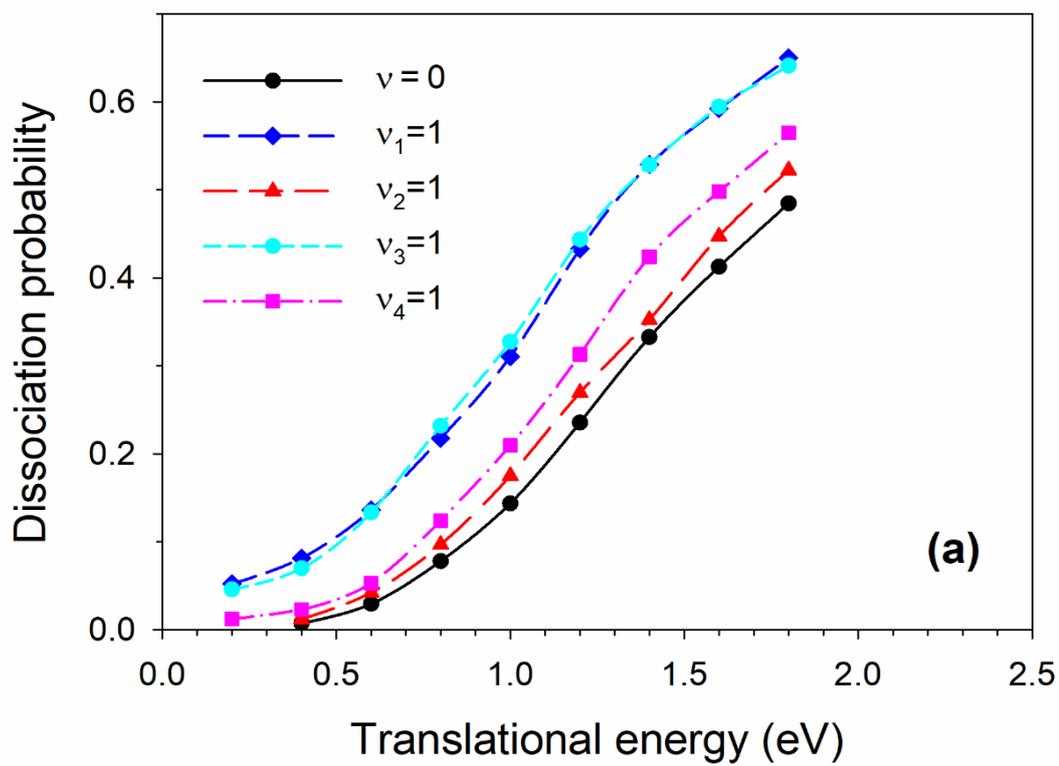

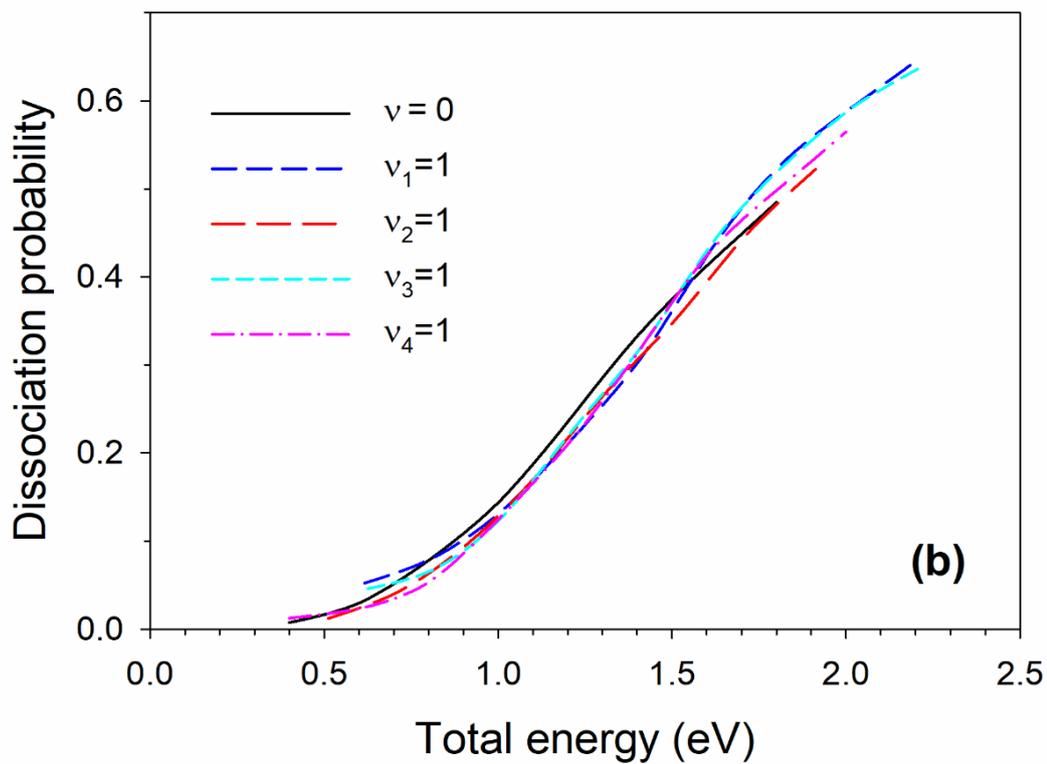

Fig. 3. Calculated dissociation probabilities of the ground and vibrational excited states of $NH_3$ as a function of (a) translational energy and (b) total energy in eV.